# Pilot Synthesis for Distributed Synchronization

Vladislav Ryzhov
Department of Multimedia Technologies and Telecommunications
Moscow Institute of Physics and Technology
Moscow, Russia
ryzhov.vp@phystech.edu

Anton Laktyushkin
Department of Multimedia Technologies and Telecommunications
Moscow Institute of Physics and Technology,
Moscow, Russia
laktyushkin@yahoo.com

*Abstract* — Multiuser Massive MIMO is ongoing wireless technology with high prospects in the distributed architecture with various system design and splitting options. Any of such solutions suffers from imperfect synchronization that can be reduced via promising Over the Air pilot exchanges. In this paper we introduce requirements and algorithm for flexible non orthogonal pilot design for time synchronization without changes to 3GPP protocol stack. Our solution decreases number of synchronization slots and solves problem of separating guard intervals.

## I. INTRODUCTION

Distributed MIMO technology provides spectral efficiency gain by increase of channel rank [1]. Unfortunately, such system requires more accurate synchronization [2] to achieve an increase in throughput. To be implemented into the product, the new synchronization solution must be scalable and meet the current system requirements. The article [3] introduces scalable synchronization and calibration principles. In paper [4] authors proposed distributed time synchronization algorithm. Multiple exchanges among all nodes have to be done in order to overcome impairments with Least Squares (LS) estimator [5] (authors of [6] analyzed cooperative network synchronization). But in dense urban scenarios spatial separation is sometimes insufficient that causes long-lasting sequential synchronization and performance reduction. Also, Inter Symbol Interference (ISI) cancellation issue (caused by initial rough network time synchronization, hardware mismatches and residual Time Alignment Errors (TAE)) requires insertion of guard intervals reducing capacity as well.

Current synchronization solution is based on point-to-point (P2P) master-slave synchronization while consensus algorithms [7] perform better for dense networks. For rough stage Primary, Secondary and Resynchronization Synchronization Signals [8] (PSS, SSS and RSS) are processed is downlink (DL). As defined in [9]-[10] they are generated from Zadoff-Chu or m-sequences (authors of [11] analyzed performance of 5G time synchronization) which mainly exploit the only characteristic – good autocorrelation properties. Additionally, each physical channel contains specific pilots (e.g., Demodulation Reference Signal for Physical Uplink / Downlink Shared Channels (PUSCH/PDSCH)), that might be used for fine time synchronization. Design of these pilots may be improved for distributed architecture as it requires perfect synchronization between Access Points (AP) as well as between APs and User Equipment (UE). Distributed synchronization pilots may be introduced in DL or uplink (UL) physical channels at the same slots depending on the receiver type and cell operation mode.

This paper considers issues related to pilot design for time synchronization for arbitrary system with distributed architecture and mesh/tree alignment algorithms. Related problems are solved via proposed pilot synthesis algorithm with an emphasis on the limitation of commercial system.

## II. SYSTEM ARCHITECTURE

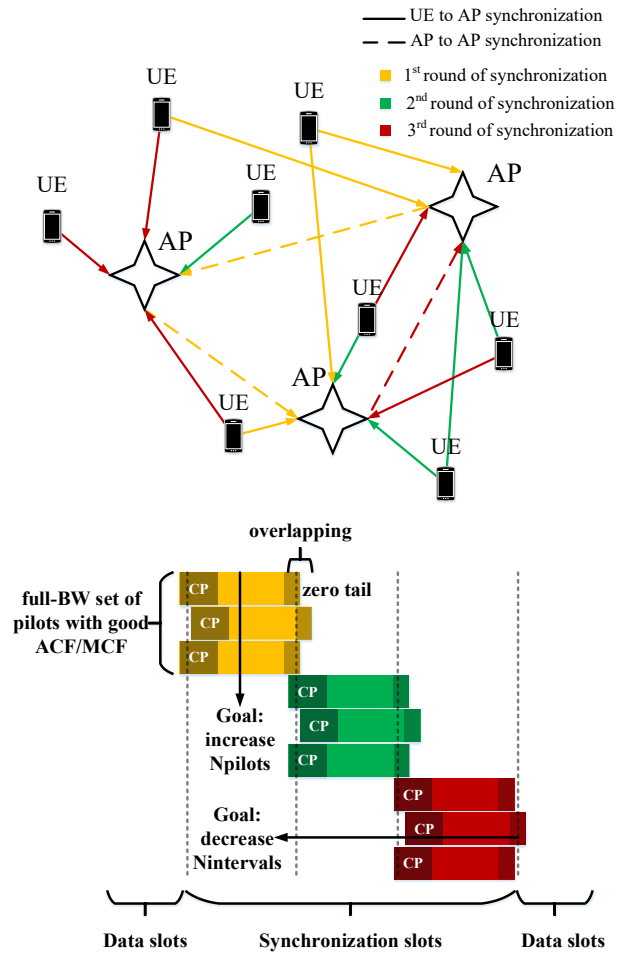

Fig. 1. Overall representation of synchronization process.

This paper considers arbitrary Radio Access Network (RAN) architecture that requires accurate timing synchronization via Over the Air (OTA) exchanges between many nodes as shown in Fig. 1. APs and UEs are located in typical urban environment with multipath propagation. Hardware time mismatches, two-way First Arrival Path (FAP) detection errors, variable wireless channel conditions cause symbol overlapping. Clock difference measurements, pilots' exchange order, clock alignment procedure, etc. go beyond the scope of this paper.

Accurate LS clock difference estimation requires measurement across many pairs of nodes according to some schedule. Specific function in multi-RAT protocol stack searches for optimal order of exchanges for synchronization. And Central Unit (CU) notifies each Distributed Unit (DU) via F1-C interface about estimated schedule. Multiple pairs of APs and UEs simultaneously perform P2P or multicast pilot exchanges for several sequential OFDM symbols according to received schedule.

This paper follows the trend of substitution of components of expensive wearable devices, smartphones and base stations with cheaper ones with efficient algorithms and considers mesh time synchronization procedure without any anchor nodes. Such solution is robust to hardware impairments and varying conditions. RMSE of time alignment for joint estimation with multiple measurements is less than for single P2P estimation even for precise devices. This emphasizes the need for simple solution with multiple simultaneous measurements. Another reason is synchronization of many wireless devices that can be done only through dedicated anchors or with exploiting additional measurements without increase of allocated resources. The latter requires simultaneous measurement of maximum possible number of pairs of transceivers.

## III. TECHNICAL REQUIREMENTS

Design of synchrosignals for distributed system is limited by the following commercial system restrictions:

### A. Compliance with OFDM

In order to minimize complexity of time synchronization procedure pilot design should satisfy general principles of OFDM and data exchange between layers in 3GPP protocol stack. Therefore, pilots shall be inserted in frequency domain (FD) limited by allocated bandwidth without access to time domain (TD).

### B. Relatively good Autocorrelation and Mutual Correlation function (ACF and MCF) properties

Designed pilots have to correspond to maximum MainPeak-to-SidePeak value of MCF inside search window which is limited by initial time synchronization (e.g., IEEE 1588 protocol accuracy is less than 1μsec[12]), delays, residual errors, etc.

### C. Time efficient

Synchronization process must utilize minimum time resources that requires minimizing the number of synchronization slots provided by maximization the number of simultaneous transmissions.

### D. Robust to overlapping

Designed pilots should not cause ISI during synchronization process or worsen data transmission in case of overlapping (even in case with many simultaneous transmissions).

### E. Flexible

Synchrosignal has to be flexible regarding number of nodes, number of simultaneous transmissions, requirements for initial and residual TAE, overlapping length, accuracy of peak detection, etc.

Listed restrictions limit the use in TD of classic sequences (e.g., gold sequences) or computationally designed [13]-[14] with high cross-correlation. In order to synthesize such sequences in FD without ISI bandwidth has to be limited, which causes symmetric overlapping of impulse response due to FFT properties (idea of algorithm [13] with search of pilots in orthogonal projection to correlation matrix is inappropriate, because orthogonal projection includes pilots with non-zero TD tail). The main problem to tackle is to manage TD behavior of pilots from FD in presence of singular mapping.

## IV. PILOT SYNTHESIS

### A. Pilot design principle

Consider the task of FD pilot synthesis with $N_{sc}$ subcarriers, $N_{fft}$ – FFT length, T – number of zero samples in TD, W – IFFT matrix.

$$W = \begin{bmatrix} W_1^{N_{fft} \times N_{sc}} & W_2^{N_{fft} \times (N_{fft} - N_{sc})} \end{bmatrix} \quad (1)$$

$$W = \begin{bmatrix} W_{11}^{(N_{fft}-T) \times N_{sc}} & W_{12}^{(N_{fft}-T) \times (N_{fft}-N_{sc})} \\ W_{21}^{T \times N_{sc}} & W_{22}^{T \times (N_{fft}-N_{sc})} \end{bmatrix} \quad (2)$$

Performing Singular Value Decomposition (SVD) of W submatrix we find matrix of right eigenvectors $V_0$:

$$W_{21} = USV^H, V = \begin{bmatrix} V_1^{N_{sc} \times T} & V_0^{N_{sc} \times (N_{sc}-T)} \end{bmatrix} \quad (3)$$

$V_0$ – Linear operator, mapping any FD pilot preimage x to FD pilot $y_{FD}$ with zero tail in TD:

$$y_{FD} = V_0 x, \forall x \in \mathbb{C}^{N_{sc}-T} \quad (4)$$

The aim of algorithm is to find preimages $\{x_n\}_{n=1}^{N_{pilots}}$ that correspond to FD pilots $\{y_n\}_{n=1}^{N_{pilots}}$ and TD pilots $\{Ax_n\}_{n=1}^{N_{pilots}}$ with good ACF/MCF properties.

### B. Algorithm

Cost function consists of several components:

$$F(x, n) = F_1(x, n) + F_2(x, n) + F_3(x, n) \quad (5)$$

where x – preimage of pilot, n – number of samples to shift.

- ACF cost function

$$F_1(x, n) = \frac{[c_n(x^H A^H)(Ax)][(x^H A^H)r_n(Ax)]}{[(x^H A^H)(Ax)]^2} \quad (6)$$

- MCF cost function

$$F_2(x, n) = \sum_{i=1}^{N_{pilots}} \frac{[c_n(x^H A^H)(Ax_i)][(x_i^H A^H)r_n(Ax)]}{[(x^H A^H)(Ax)]^2} \quad (7)$$

$A = W_1 V_0$ – Linear operator[1]: pilot preimages → TD pilot.
$c_n/r_n$ – operator of cyclic shift of columns/rows in matrix or vector.

$F_3(x, n)$ corresponds to external constraints (e.g., requirement for low-PAPR pilot). Refer to appendix for details.

Min-max optimization problem is treated as search of zero-tail pilot preimage with smallest side peak in MCF:

$$x = \underset{x}{\operatorname{argmin}} \left\{ \max_{n,i}\bigl(F_i(x, n)\bigr) \right\} \quad (8)$$

can be solved via two gradient descend methods:

$$\vec{g}_1(x) = \frac{\partial F_{i_{max}}(x, n_{max})}{\partial x} \quad (9)$$

$$\vec{g}_2(x) = \frac{\partial \left[ \sum_i \alpha_i \sum_{k=1}^{N_{peaks}} \beta_k \tilde{F}_i(x, k) \right]}{\partial x} \quad (10)$$

$$\tilde{F}_i(x, n) = \underset{n}{\operatorname{sorted}}\bigl(F_i(x, n)\bigr) \quad (11)$$

The former involves finding the only greatest side peak across all cost function components and descend to its minimum. The latter method requires descending to minimum of several weighted peaks of all cost function components.

---

**Pilot search algorithm**

1) *Initialization*
   $x_1 = x_0$
2) *Find gradient step*
   Across all components $F_i$ find $N_{peaks}$ maximum peaks and matching gradients.
   
   $$\begin{pmatrix} n_k^{max} \in \left[ \frac{-T_{max}}{2}, \frac{T_{max}}{2} \right] | \\ n_k^{max} \notin \left[ \frac{-T_{min}}{2}, \frac{T_{min}}{2} \right], i = 1 \end{pmatrix}$$
   
   Find gradient step:
   $$\vec{g}(x) = \sum_{i, \{n_k\}_{k=1}^{N_{peaks}}} \alpha_i \beta_k \frac{\partial F_i(x, n_k)}{\partial x}$$
3) *Update pilot preimage*
   $\vec{x}_n = \vec{x}_{n-1} - h_{step}^{(n)} \cdot \vec{g^H}(x_{n-1})$
4) *Repeat steps 2-3 for all pilots*
5) *End condition*
   if $\|x_n - x_{n-1}\| < \varepsilon$:
       $y_{FD} = V_0 x_n$
       end.
   else:
       go to step 2.

---

Fig. 2 contains idea of an algorithm: for current pilot in set, we find time shifts that give $N_{peaks}$ maximum undesirable peaks in a bounded region and make gradient step to minimize them. The same should be repeated for other pilots. During next iteration we find new peaks and make gradient step towards its minimum. Such alternating pilots and peaks procedure with decreasing step size $h_{step}^{(n)}$ converges to desired synchrosignals.

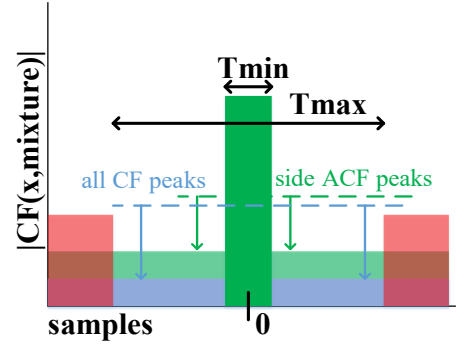

Fig. 2. Visualization of algorithm idea.

### C. Compliance with the requirements

Pilots constructed above could be integrated in OFDM symbol with any dedicated carrier order. Proposed method allows to add any optimization criteria according to external constraints. Thanks to the way algorithm designed it is possible to find arbitrary combinations of pilots for current operating conditions:

- MainPeak-to-SidePeak ratio may be increased via $T_{max} \to min$, $T_{min} \to max$, $T \to min$ with more degrees of freedom for pilots.

- Interval $T_{min}$ influences accuracy of clock difference estimation. For first synchronization slots $T_{min}$ can be increased to perform rough time alignment and for next slots procedure may be repeated with narrower $T_{min}$. Also, for fast varying channels with strong multipath large FAP detection error constrains narrowest $T_{min}$.

- Interval $T_{max}$ limits maximum clock difference between two nodes. For cheap oscillators $T_{max}$ interval should be long enough, but for temperature- or oven-controlled crystal oscillators $T_{max}$ may correspond to narrower clock difference search area.

- Interval T can be selected based on ISD, propagation environment (multipath, delay spread), residual hardware delays mismatches to keep tradeoff between ISI and MainPeak-to-SidePeak ratio.

## V. SIMULATION RESULTS

Table 1 contains input parameters that were used for pilot synthesis.

TABLE I. SIMULATION PARAMETERS

| Parameter | Value | Comments |
|---|---|---|
| $N_{pilots}$ | 4 | --- |
| $N_{fft}$ | 4096 | --- |
| $N_{sc}$ | 3300 | $\Delta f = 30 kHz, BW = 99 MHz$ |
| T | 1750 | Zero tail: $\frac{T}{\Delta f \cdot N_{fft}} \sim 14\ \mu s$ |

---

[1] $A$ – is singular, notation $A^{-1}$ – treat as approximate LS inversion via SVD with discard of zero diagonal elements.

| Parameter | Value | Comments |
|---|---|---|
| $T_{min}$ | 2 | Precision: $\frac{1+T_{min}}{2\Delta f \cdot N_{fft}} \sim 12$ ns |
| $T_{max}$ | 370 | Initial clock difference: $\frac{T_{max}}{2\Delta f \cdot N_{fft}} \sim 1.5$ μs |

Pilots were initialized with complex random numbers. Fig. 3 represents convergence of algorithm with the gradient descend for the only maximum side peak.

The following strategies for gradient descend were analyzed:

1) *Descend method*
   a) Maximum peak gradient
   b) Lagrange-like (weighted gradients)
2) *Gradient averaging*

$$\overrightarrow{g^{(n)}}(x) = \alpha_{learn\ rate}\vec{g}(x) + (1 - \alpha_{learn\ rate})\overrightarrow{g^{(n-1)}}(x)$$

3) *Strategy for choice of step size*

   a) $h_{step}^{(n)} = \frac{h_{step}^{(n-1)}}{a} \mid F^{(n)} > F^{(n-1)}$
   
   b) $h_{step}^{(n)} = f(n)$
   
   c) $h_{step}^{(n)} = f(F^{(n)})$

4) *Number of peaks in ACF/MCF*

The fastest option for min-max problem was one with direct update of gradient vector for the single maximum peak due to extra computations for additional gradients. For larger sets of pilots gradient averaging improves convergence. The fastest option for step size was to choose it according to the value of cost function.

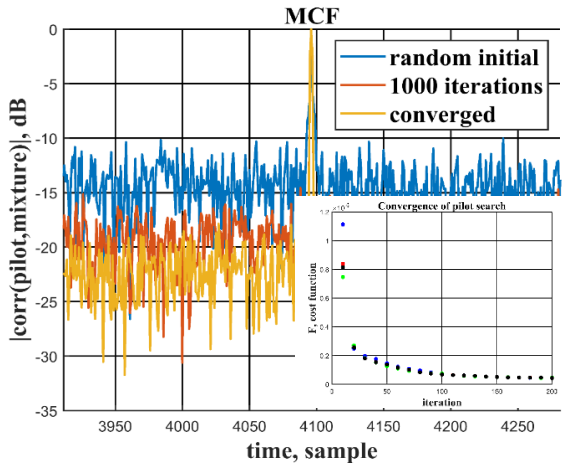

Fig. 3. Algorithm convergence for 4 multiplexed pilots.

Fig. 4 demonstrates 1 pilot out of converged set of 4 pilots in FD/TD. Zero tail in TD is provided by operator $V_0$. Fig. 5 shows that required ACF/MCF properties were satisfied.

The worst case for distributed synchronization is depicted in Fig. 6. It corresponds to system[2] with 4 pairs of transceivers that simultaneously exchange set of 4 precomputed pilots. Path Loss (PL) values of main (synchronization) link and interfered (adverse parallel synchronization among other devices) links are the same (this case should be excluded via correct scheduling). Even in such complex conditions designed pilots demonstrate performance very close to MCF peak value.

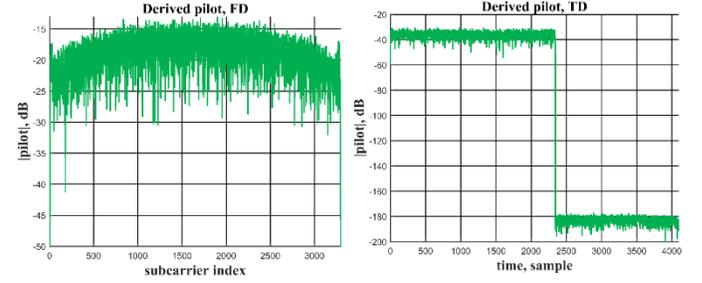

Fig. 4. Representation of 1 of converged set of 4 pilots in TD/FD.

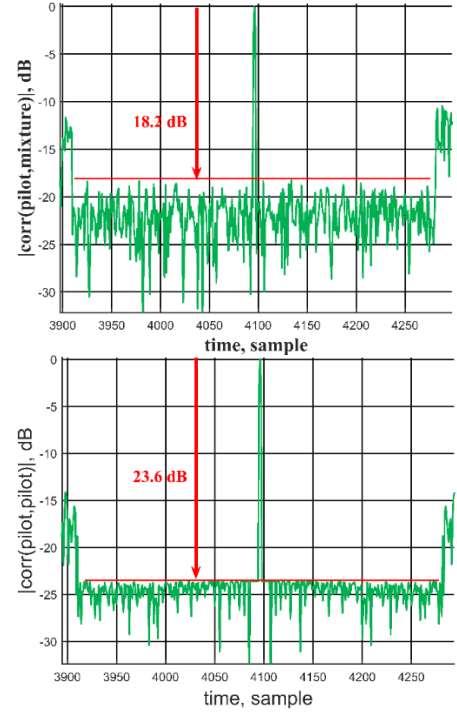

Fig. 5. ACF/MCF of converged set of 4 pilots.

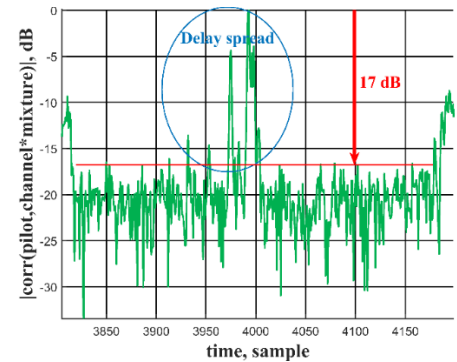

Fig. 6. MCF of converged set of 4 pilots passed through channels.

If number of devices is large enough, they need to perform more simultaneous transmissions due to the limitation of the number of synchronization slots. Table 2 represents pilot synthesis results for various size of pilot sets. MCF peak value was analyzed for the worst case (equal power links). Correct scheduling should take into account connectivity table filled with PL values. In this case MainPeak-to-SidePeak value will be closer to ACF peak value, than MCF peak value. For large number of simultaneous transmissions improvement of gradient search is another question that has to be studied as improvement of alternating procedure for ACF/MCF peaks and pilots. Table 2 also demonstrates that number of synchronization slots may be decreased with proposed scheme. In the basic solution, 3 pairs of devices simultaneously exchange pilots, each of which occupies a quarter of the band. In this case zero tail will be of length:

$$\left(3300 \text{ subcarriers} \cdot \left[1 - \frac{3}{4}\right] \cdot \frac{1}{\Delta f N_{fft}} = 6.7 \text{ μsec}\right)$$

---

[2] Channels were modeled via C++ Ray Tracing simulator in Urban Street environment.

TABLE II. MULTIPLEXING SYNCHROSIGNALS

| N pilots=4 | N pilots=8 | N pilots=64 |
|---|---|---|
| 10log(|corr (pilot, mixture)|), MainPeak-to-SidePeak, dB | | |
| 18.2 dB | 15 dB | 8 dB |
| number of synchronization slots | | |
| -25% | -62.5% | -95.3% |

## VI. CONCLUSIONS

This paper outlined and solved the problem of pilot synthesis for synchronization for systems with distributed architecture. Proposed solution can be implemented in commercial system with a high flexibility making possible to add extra optimization criteria and select the precomputed set of pilots online for current conditions. Moreover, proposed approach may include quasistatic channels for better performance while computing sets of pilots in remote server. Our procedure reduces allocated time slots for synchronization and solves problem of pilots overlapping.

8 pairs of transceivers may exchange synchrosignals providing 15 dB MainPeak-to-SidePeak value. With increase of the number of devices up to 64 pairs may exchange simultaneously with 8 dB main peak (the worst case that can be enhanced with correct scheduling/power matching) and 20x less allocated slots.

Some of questions not disclosed include: upper bound of MainPeak-to-SidePeak value of designed pilots under the given restrictions and improvement of convergence for large sets of pilots.

## APPENDIX

### A. Low-PAPR constraint for pilot design

Additional cost function component:

$$F_3(x) = \frac{\max \text{element}(|Ax|^2)}{\text{mean element}(|Ax|^2)} \quad (12)$$

PAPR reduction idea is the same as for ACF/MCF gradient descend. The main difference is that peaks and gradients are regarded in TD and then converted back to FD.

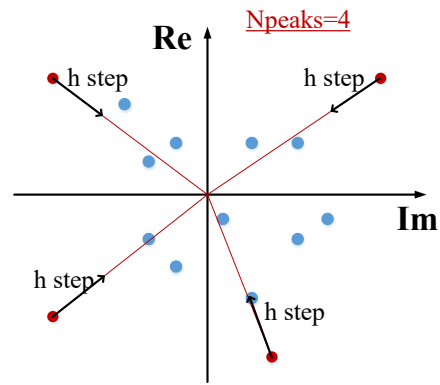

Fig. 7. Representation of TD gradient for PAPR reduction.

After ACF and MCF gradients are found, new pilot is transferred to TD and TD gradient step towards minimum of $N_{peaks}$ peaks are estimated. TD gradient step (example is shown in Fig. 7) is transferred to FD gradient via LS inverse of singular A and FD pilot is being updated.

---

**Low-PAPR pilot design procedure**

**for i=(1,N_PAPR_reductions)**
1) *Find gradient step in TD*
   Find $N_{peaks}$ maximum peaks of TD pilot and matching gradient in TD:
   $y_{TD} = Ax_{n-1}$
   $\vec{g}_{PAPR}^{TD}(x) = h_{step} \cdot |y_{TD,k}| \cdot I\{|y_{TD,k}| > \varepsilon\}$,
   
   E.g.: $\vec{g}_{PAPR}^{TD} = \begin{pmatrix} 0 \\ \ldots \\ 0 \\ |y_{TD,k}| \\ 0 \\ \ldots \\ |y_{TD,k_N}| \end{pmatrix}$

2) *Find gradient step in FD*
   $\vec{g}_{PAPR}^{FD}(x) = A^{-1}\vec{g}_{PAPR}^{TD}(x)$
3) *Update pilot preimage*
   $\vec{x}_n = \vec{x}_{n-1} - h_{step} \cdot \vec{g}_{PAPR}^{FD}(x)$

**end**

---

Table 3 contains performance analysis results of pilot synthesis algorithm with PAPR reduction technique for various ratio between MainPeak-to-SidePeak and PAPR values. Such procedure works efficiently, but slightly decrease ACF/MCF properties of designed pilots.

TABLE III. PAPR REDUCTION PERFORMANCE

| |corr (pilot, mixture)|, MainPeak-to-SidePeak, dB | | | |
|---|---|---|---|
| 18.2 dB | 17.3 dB | 16.8 dB | 16.5 dB |
| **no low-PAPR criteria** | **low-PAPR pilots** | | |
| 9 dB | 8.3 dB | 8 dB | 7.5 dB |

*B. ACF gradient*

$$\frac{\partial F_1(x)}{\partial x} = \frac{[c_n(x^H A^H)A][(x^H A^H)r_n(Ax)]}{[(x^H A^H)(Ax)]^2} + \\ + \frac{[c_n(x^H A^H)(Ax)][(x^H A^H)r_n(A)]}{[(x^H A^H)(Ax)]^2} - \\ - 2\frac{[r_n(x^H A^H)(Ax)][(x^H A^H)c_n(Ax)]}{[(x^H A^H)(Ax)]^3}(x^H A^H A) \quad (13)$$

*C. MCF gradient*

$$\frac{\partial F_2(x)}{\partial x} = \frac{[c_n(x^H A^H)(Ax_i)][(x_i^H A^H)r_n(A)]}{[(x^H A^H)(Ax)]^2} - \\ - 2\frac{[r_n(x^H A^H)(Ax_i)][(x_i^H A^H)c_n(Ax)]}{[(x^H A^H)(Ax)]^3}(x^H A^H A) \quad (14)$$